\documentclass[12pt,manuscript,a4paper]{aastex}
\usepackage{subfigure}
\usepackage{natbib}
\bibpunct{[}{]}{,}{n}{}{;}

\begin{document}
\title{Reverberation Mapping of the Optical Continua of 57 MACHO Quasars}
\author{Justin Lovegrove \altaffilmark{1,2}}
\altaffiltext{1}{School of Physics and Astronomy, University of Southampton,
Southampton, UK SO171BJ}
\altaffiltext{2}{Harvard-Smithsonian Centre for Astrophysics, Cambridge,
Massachusetts, USA 02144}
\email{jl805@soton.ac.uk}
\tableofcontents
\newpage
\listoffigures
\newpage
\section{Abstract}
Autocorrelation analyses of the optical continua of 57 of the 59
MACHO quasars reveal structure at proper time lags of $544 \pm 5.2$
days with a standard deviation of 77 light days.  Interpreted in the context of
reverberation from elliptical outflow winds as proposed by Elvis
(2000) \cite{E00}, this implies an approximate characteristic size scale
for winds in the MACHO quasars of $544 \pm 5.2$ light days.  The
internal structure variable of these reflecting outflow surfaces is found to be
$11.87^o \pm 0.40^o$ with a standard deviation of $2.03^o$.
\newpage
\section{Introduction}
Brightness fluctuations of the UV-optical continuum emission of
quasars were recognised shortly after the initial discovery of the objects in
the 1960s \cite{ms}.  Although several programmes were
undertaken to monitor these fluctuations, little is yet known about their
nature or origin.  A large number of these have focused on comparison of the
optical variability with that in other wavebands and less on long-timescale,
high temporal resolution optical monitoring.  Many studies have searched for
oscillations on the $\sim$ day timescale in an attempt to constrain the inner
structure size (eg \cite{wm}).  This report, however, is concerned with
variability on the year timescale to evidence global quasar structure.

In a model proposed by Elvis \cite{E00} to unite the various spectroscopic
features associated with different ``types'' of quasars and AGN (eg broad
absorption lines, X-ray/UV warm absorbers, broad blue-shifted emission lines),
the object's outer accretion disc has a pair of bi-conical extended narrow
emission line regions in the form of outflowing ionised winds.  Absorption and
emission lines and the so-called warm absorbers result from orientation effects
in observing these outflowing winds.  Supporting evidence for this is
provided by a correlation between polarisation and broad absorption lines found
by \cite{O}.  Outflowing accretion disc winds are widely considered to be
a strong candidate for the cause of feedback (for a discussion of the currently
understood properties of feedback see \cite{FEA}).  Several models
have been developed \cite{p00,p08}
to simulate these winds.  \cite{p00} discusses different launch
mechanisms for the winds - specifically the balance between magnetic forces and
radiation pressure - but finds no preference for one or the other, while
\cite{p08} discusses the effect of rotation and
finds that a rotating wind has a larger thermal energy flux and lower mass
flux, making a strong case for these winds as the source of feedback.  The
outflow described by \cite{E00} is now usually identified with the
observationally-invoked ``dusty torus'' around AGN \cite{ra}.

\cite{mp} demonstrated for the MACHO quasars that there is no
detectable lag time between the V and R variability in quasars, which can be
interpreted as demonstrating that all of the optical continuum variability
originates in the same region of the quasar.

\cite{ST97} observed the gravitationally lensed quasar Q0957+561 to
measure the time delay between the two images and measure microlensing effects.
In doing so, they found a series of autocorrelation subpeaks initially
attributed to either microlensing or accretion disc structure.  These results
were then re-interpreted by \cite{S05} as Elvis' outflowing winds at a
distance of $2 \times 10^{17} cm$ from the quasar's central compact object.  A
model applied by \cite{VA07} to the quasar Q2237+0305, to simulate
microlensing, found that the optimal solution for the system was one with a
central bright source and an extended structure with double the total
luminosity of the central source, though the outer structure has a lower
surface brightness as the luminosity is emanating from a larger source, later
determined by \cite{slr08} to lie at $8.4 \times 10^{17} cm$.

\cite{slr08}
continued on to argue that since magnetic fields can cause both jets and
outflows, they therefore must be the dominant effect in AGN.  \cite{lslp}
however pointed out that the magnetic field required to power the observed
Elvis outflows is too great to be due to the accretion disc alone.  They
therefore argue that all quasars and AGN have an intrinsically magnetic central
compact
object, which they refer to as a MECO, as proposed by \cite{rl07}, based upon
solutions of the Einstein-Maxwell equations by \cite{rl03}.  One compelling
aspect to this argument is that it predicts a power-law relationship between
Elvis outflow radius and luminosity, which was found in work by Kaspi et
al \cite{ks} and updated by Bentz et al \cite{b}, if one assumes the source
of quasar broad emission lines to be outflow winds powered by magnetic fields.
The \cite{ks} and \cite{b} results were in fact empirically derived for AGN
of $Z < 0.3$ and \cite{ks} postulates that there may be some evolution of this
relation with luminosity (and indeed one might expect some time-evolution of
quasar properties which may further modify this scaling relation) so
generalising these results to quasars may yet prove a fallacy.  The radius of
the broad line region was found to scale initially by \cite{ks} as $R_{blr}
\propto L^{0.67}$, while \cite{b} found $R_{blr} \propto L^{0.52}$.

Another strength of the MECO argument is that while \cite{wu} found quasar
properties to be uncorrelated using the current standard black hole models,
\cite{lslp} and \cite{sll} found a homogeneous population of quasars using the
\cite{rl03} model.  \cite{p} used microlensing observations of 10
quadruply-lensed quasars, 9 of which were of known redshift including
Q2237+0305, to demonstrate that standard thin accretion disc models, such as
the widely-accepted Shakura-Sunyaev (S-S) disc \cite{ss73}, underestimate the
optical continuum emission region thickness by a factor of between 3 and 30,
finding an average calculated thickness of $3.6 \times 10^{15} cm$, while
observed values average $5.3 \times 10^{16} cm$.  \cite{bp} found a radius of
the broad line region for the Seyfert galaxy NGC 5548 of just under 13 light
days when the average of several spectral line reverberations were taken,
corresponding to $R_{blr} = 3.3 \times 10^{16} cm$.  When the scaling of
\cite{ks} and \cite{b} is taken into account, the \cite{bp} result is
comparable to the \cite{slr08} and \cite{S05} results (assuming
$\frac{L_{quasar}}{L_{seyfert}} \sim 10^4$, then \cite{b} would predict a
quasar $R_{blr}$ of approximately $3 \times 10^{18}$).  Also, given that black
hole radius scales linearly with mass, as does the predicted radius of the
inner edge of the accretion disc, a linear mass-$R_{blr}$ relationship might
also be expected.  Given calculated Seyfert galaxy black hole masses of order
$10^8M_o$ and average quasar masses of order $10^9M_o$, this would scale the
Seyfert galaxy $R_{blr}$ up to $3.3 \times 10^{17} cm$.  While these relations
are not self-consistent, either of them may be found consistent with the
existing quasar structure sizes.  \cite{r} also found
structure on size scales of $10^{16} cm$ from microlensing of SDSS J1004+4112
which would then scale to $10^{18}cm$.  These studies combined strongly
evidence the presence of the Elvis outflow at a radial distance of
approximately $10^{18} cm$ from the central source in quasars which may be
detected by their reverberation of the optical continuum of the central quasar
source.

The \cite{VA07} result is also in direct conflict with the S-S accretion disc
model, which has been applied in several unsuccessful attempts to describe
microlensing observations of Q2237.  First a simulation by \cite{w} used the
S-S disc to model the microlensing observations but predicted a large
microlensing event that was later observed not to have occurred.  \cite{k} then
attempted to apply the S-S disc in a new simulation but another failed
prediction of large-amplitude microlensing resulted.  Another attempt to
simulate the Q2237 light curve by \cite{ei} produced the same large-amplitude
microlensing events.  These events are an inherent property of the S-S disc
model where all of the luminosity emanates from the accretion disc, hence
causing it all to be lensed simultaneously.  Only by separating the luminosity
into multiple regions, eg two regions, one inner and one outer, as in
\cite{VA07}, can these erroneous large-amplitude microlensing events be
avoided.

Previous attempts have been made to identify structure on the year timescale,
including structure function analysis by Trevese et al \cite{t} and by Hawkins
\cite{h96,h06}.  \cite{t} found strong anticorrelation on the $\sim 5$ year
timescale but no finer structure - this is unsurprising as their results were
an average of the results for multiple quasars taken at low temporal
resolution, whereas the size scales of the Elvis outflow winds should be
dependant on various quasar properties which differ depending on the launch
mechanism and also should be noticed on smaller timescales than their
observations were sensitive to.  \cite{h96} also found variations on the $\sim
5$-year timescale again with poor temporal resolution but then put forth the
argument that the variation was found to be redshift-independant and therefore
was most likely caused by gravitational microlensing.  However \cite{cs}
demonstrated that microlensing occurs on much shorter timescales and at much
lower luminosity amplitudes than these long-term variations.  \cite{h06} used
structure function analysis to infer size scales for quasar accretion discs but
again encountered the problem of too infrequent observations.  In this paper
and \cite{h07} it was argued that Fourier power specra are of more use in the
study of quasars, which were then used in \cite{h07} to interpret quasar
variability.  However, since reverberation is not expected to be periodic,
Fourier techniques are not suited to its detection.  Hawkins' observations of
long-timescale variability were recognised in \cite{slp} as a separate
phenomenon to the reverberation expected from the Elvis model; this long-term
variability remains as-yet unexplained.  Hawkins' work proposed this variation
to be indicative of a timescale for accretion disc phenomena, while others
explained it simply as red noise.

\cite{u}
demonstrated that there is a correlation between the optical and x-ray
variability in some but not all AGN, arguing that x-ray reprocessing in the
accretion disc is a viable source of the observed variability, combined with
viscous processes in the disc which would cause an inherent mass-timescale
dependance, as in the S-S disc the temperature at a given radius is
proportional to mass$^{- \frac{1}{4}}$.  In this case a red-noise power
spectrum would describe all quasar variability.  For the purposes of this
report however, the S-S disc model is regarded as being disproven by the
\cite{VA07} and \cite{p} results in favour of the Elvis outflow model and so
the temperature-mass relation is disregarded.  A later investigation \cite{as}
demonstrated that the correlation between X-ray and optical variability cannot
be explained by simple reprocessing in an optically thick disc model with a
corona around the central object.  This lends further support to the assumption
that the \cite{VA07} model of quasars is indeed viable.

\cite{gi} demonstrated from Fourier power spectra that shorter-duration
brightness events in quasars statistically have lower amplitudes but again
their temporal resolution was too low to identify reverberation on the expected
timescales.  Also as previously discussed, non-periodic events are extremely
difficult to detect with Fourier techniques.  All of these works are biased by
the long-term variability recognised by \cite{slp} as problematic in QSO
variability study.  A survey by \cite{r04} proposed that quasars could be
identified by their variability on this timescale but the discovery by
\cite{slp} that long-timescale variation is not quite a universal property of
quasars somewhat complicates this possibility.  The spread of variability
amplitudes is also demonstrated in \cite{n} for 44 quasars observed at the Wise
Observatory.  The \cite{n} observations are also of low temporal resolution and
uncorrected for long-term variability thus preventing the detection of
reverberation patterns.

This project therefore adopts the \cite{E00} and \cite{VA07} model of
quasars consisting of a central compact, luminous source with an accretion disc
and outflowing winds of ionised material with double the intrinsic
luminosity of the central source.  The aim is to search for said winds via
reverberation mapping.  Past investigations such as \cite{bot} have attempted
reverberation mapping of quasars but have had large gaps in their data,
primarily due to the fact that telescope time is usually allocated only for a
few days at a time but also due to seasonal dropouts.  Observations on
long timescales, lacking seasonal dropouts and with frequent
observations are required for this purpose, to which end the MACHO programme
data have been selected.  Past results would predict an average radius of the
wind region of order $10^{18} cm$.

The assumption will be made that all structure on timescales in the region of
$c \Delta t = 10^{18} cm$ is due to reverberation and that the
reverberation process is instantaneous.  The aim is simply to verify whether
this simplified version of the model is consistent with observation, not to
compare the model to other models.

\section{Theoretical Background}
Reverberation mapping is a technique whereby structure
size scales are inferred in an astrophysical object by measuring delay times
from strong brightness peaks to subsequent, lower-amplitude peaks.  The
subsequent peaks are then assumed to be reflections (or absorptions and
re-emissions) of the initial brightness feature by some external structure (eg
the dusty torus in AGN, accretion flows in compact objects).  Reverberation
mapping may be used to study simple continuum reflection or sources of specific
absorption/emission features or even sources in entirely different wavebands,
by finding the lag time between a brightness peak in the continuum and a
subpeak in the emission line or waveband of interest.  This technique has
already been successfully applied by, among many others, \cite{bp} and
\cite{S05}.

When looking for continuum reverberation one usually makes use of the
autocorrelation function, which gives the mean amplitude, in a given light
curve, at a time $t + \Delta t$ relative to the amplitude at time $t$.  The
amplitude of the autocorrelation is the product of the probability of a later
brightening event at dt with the relative amplitude of that event, rendering it
impossible to distinguish by autocorrelation alone, eg a 50\% probability of a
50\% brightening from a 100\% probability of a 25\% brightening, since both
will produce the same mean brightness profile.  The mathematical formulation of
this function is:
\begin{equation}
AC(\Delta t) = \frac{1}{N_{obs}} \cdot \sum  _t \frac{I(t + \Delta t) \cdot I(t)}{\sigma ^2}
\end{equation}
where I(t) represents the intensity at a time t relative to the mean intensity,
such that for a dimming event I(t) is negative.  Hence $AC(\Delta t)$ becomes
negative if a brightening event at t is followed by a dimming event at $t+
\Delta t$ or if a dimming
event at t is followed by a brightening event at $t+\Delta t$.  $\sigma$ is
the standard deviation of I - in rigorous mathematics the $\sigma ^2$ term is
in fact the product of the standard deviations for I(t) and I($t + \Delta t$)
but since they are idential in this case, $\sigma ^2$ may be used.  The nature
of the autocorrelation calculation also has a tendancy to introduce predicted
brightness peaks unrelated to the phenomena of interest.  Given autocorrelation
peaks at lag times $t_1$ and $t_2$, a third peak will also be created at lag
$t_2 - t_1$, of amplitude $\frac{A(t_2)}{A(t_1)}$, which is then divided by the
number of data points in the brightness record.

\section{Observational Details}
The MACHO survey operated from 1992 to 1999,
observing the Magellanic Clouds for gravitational lensing events by Massive
Compact Halo Objects - compact objects in the galactic halo, one of the primary
dark matter candidates.  The programme was undertaken from the Mount Stromlo
Observatory in Australia, where the Magellanic Clouds are circumpolar,
giving brightness records free from seasonal dropouts.  Information on the
equipment used in the programme can be found at http://wwwmacho.anu.edu.au/ or
in \cite{macho}.  59 quasars were within the field of view of this
programme and as a result highly sampled light curves for all of these objects
were obtained.  These quasars were observed over the duration of the programme
in both V and R filters.  The brightness records for the 59 quasars in both V
and R filters are freely available at http://www.astro.yale.edu/mgeha/MACHO/
while the entire MACHO survey data are available from 
http://wwwmacho.anu.edu.au/Data/MachoData.html.

\section{Theoretical Method}
Throughout this study the following assumptions were made:
\begin{enumerate}
\item
That all autocorrelation structure on the several hundred day timescale was
due to reverberation from the Elvis outflows, which are also the source of the
broad emission lines in quasars.  We denote the distance to this region as
$R_{blr}$ - the ``radius of the broad line region''.
\item
That the timescale for absorption and re-emission of photons was negligable
compared to the light travel time from the central source to the outflow
surface.
\item
That the internal structure variable was less than $45^o$ for all objects as
found by \cite{S05} and \cite{slr08}.  Note that an internal structure variable
of $\epsilon$ and inclination angle of $\theta$ cannot be distinguished by
autocorrelation alone from an internal structure variable of $90^o - \epsilon$
with an inclination angle of $90^o - \theta$, as demonstrated in Fig. 1.  The
assumed low $\epsilon$ is further justified by the fact that the \cite{E00}
model predicts the winds to be projected at an angle of $30^o$ from the
accretion disc plane, contraining $\epsilon \leq 30^o$ as the reverberating
surfaces must lie at lower inclinations if they are in fact part of the
outflow structure.
\item
That extinction was negligable for all objects.  Reliable extinction maps of
the Magellanic Clouds are not available and information about the quasars'
host galaxies is also unavailable so extinction calculations are not possible.
This assumption should be reasonable as any extinction would have a noticeable
impact on the colour of the quasar (given the accepted relation $3.2 \cdot
E(B-V) = A_V$ where $E(B-V)$ is the colour excess and $A_V$ is the V
absorption).
\end{enumerate}

First a series of predictions were made for the relative luminosities of the 59
quasars in the sample using their redshifts (presented in \cite{mg}) and
apparent magnitudes, combined with the mean quasar SED presented in \cite{SED}.
Two estimates were produced for these, one for the V data and one for the R
data.  This calculation was of interest as it would predict the expected ratios
of $R_{blr}$ between the objects in our sample via the \cite{ks} and \cite{b}
relations.  These relations are so far only applied to nearby AGN and so this
investigation was carried out to test their universality.  Firstly, the
apparent magnitudes were converted to flux units via the equation
\begin{equation}
m = -2.5 \cdot \log{f}
\end{equation}
where m is the apparent magnitude and f the flux.  Then the redshifts of the
objects were used to calculate their distances using the tool at \\
http://astronomy.swin.edu.au/~elenc/Calculators/redshift.php with $H_0 = 71
Mpc/km/s$.  Using this distance and the relation
\begin{equation}
f = \frac{L}{4 \pi d^2}
\end{equation}
where L is the luminosity, f the flux and d the distance, the luminosity of
each source in the observed waveband was calculated.  It was
recognised, however, that simply taking the ratio between these luminosities
was not a fair representation of the ratio of their absolute luminosities as
cosmological redshift would cause the observed region of the quasar spectrum
to shift with distance and the luminosity varies over the range of this
spectrum.  Hence the mean quasar SED of \cite{SED} was used to convert the
observed luminosities of the objects to expected luminosities at a common
frequency - in this case 50 GHz.  These values were then divided by the minimum
calculated luminosity so that their relative values are presented in Table 1.
For this purpose it was assumed that the spectral shape of the quasar is
independant of bolometric luminosity.

Next a calculation was made to predict the dependance
of the reverberation pattern of a quasar on its orientation to the observer's
line of sight.  This was produced using the geometric equations initially
presented in \cite{S05} but reformulated to become more general.  \cite{S05}
discusses "case 1" and "case 2" quasars with different orientations -
"case 1" being where the nearest two outflow surfaces lie on the near side of
the accretion disc plane and "case 2" being where the nearest two surfaces
are on the near side of the rotation axis.  These equations become generalised
by recognising that the two cases are in fact degenerate and from reverberation
alone one cannot distinguish a $13^o$ internal structure variable in "case
1" from a $77^o$ internal structure variable in "case 2".  This is also
demonstrated in Fig. 1.  For the purposes of Fig. 1, the \cite{S05}
interpretation of $\epsilon$ is adopted but later it will be demonstrated that
in fact there are several possible meanings of $\epsilon$, though this
degeneracy is the same for all interpretations.

\begin{figure}
\centering
\subfigure[\textbf{$\epsilon$ and $\theta$ as in Case I of \cite{S05}}]{\includegraphics[scale=0.5]{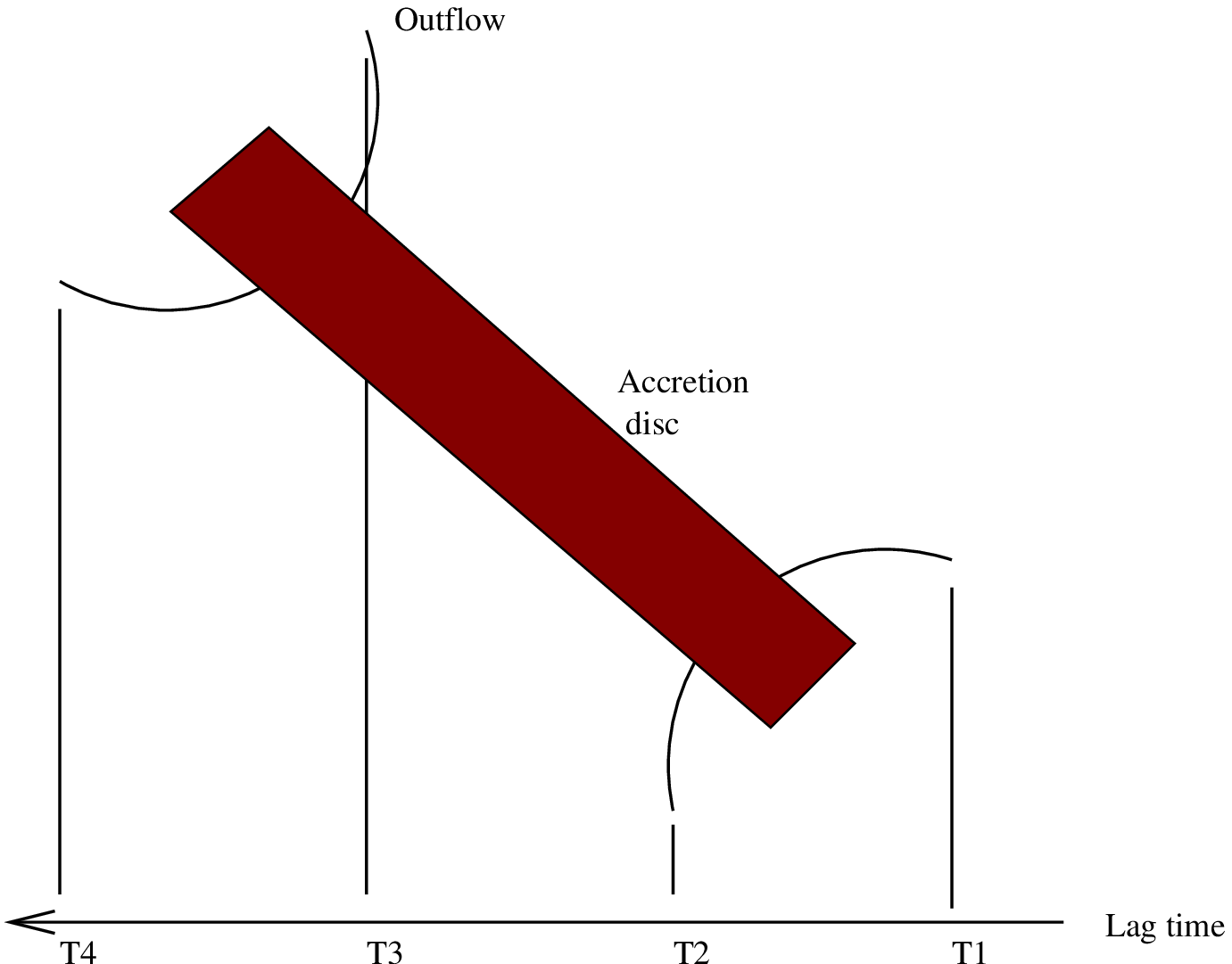}}
\subfigure[\textbf{$\theta^1 = \epsilon$ and $\epsilon^1 = \theta$}]{\includegraphics[scale=0.5]{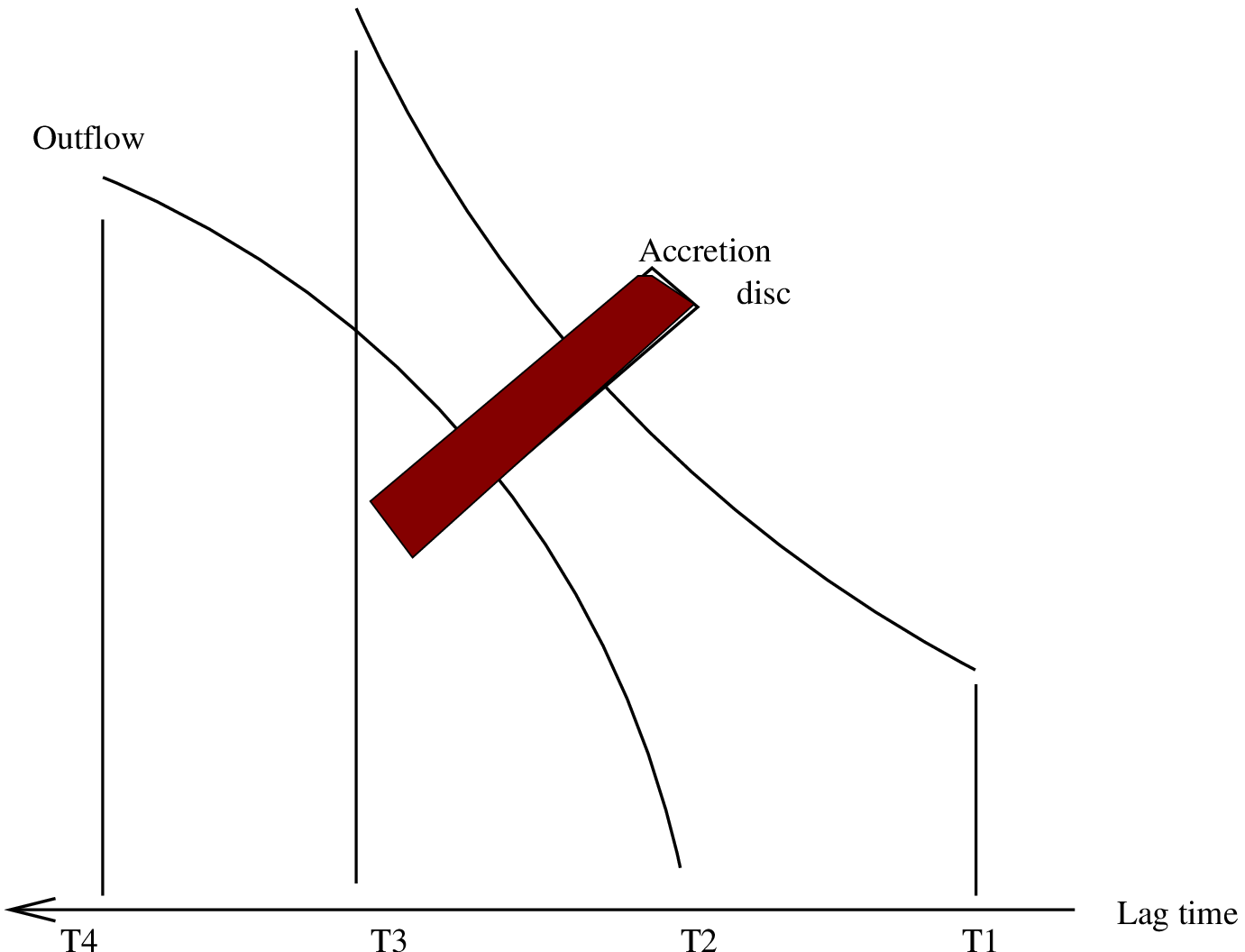}}
\subfigure[\textbf{$\theta^1 = 90^o - \theta$ and $\epsilon^1 = 90^o - \epsilon$}]{\includegraphics[scale=0.5]{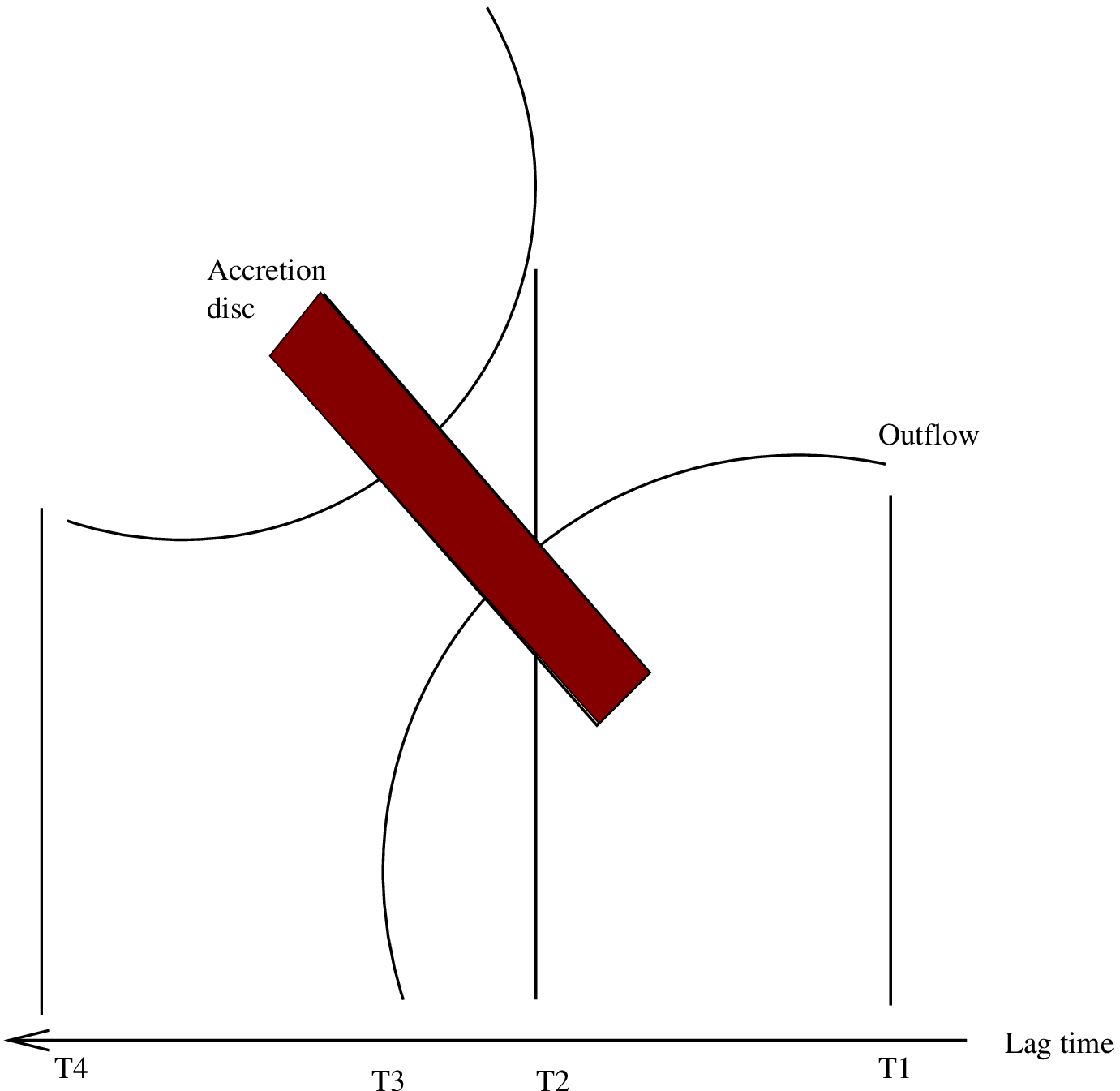}}
\caption[Degeneracy in $\theta$ and $\epsilon$]{Three different quasar
orientations and internal structure variables can give the same series of
reverberation lags}
\end{figure}

The generalisation is then
\begin{equation}
t_1 = \frac{R_{blr}}{c} \cdot (1 - \cos(\theta - \epsilon))
\end{equation}
\begin{equation}
t_2 = \frac{R_{blr}}{c} \cdot (1 - \cos(\theta + \epsilon))
\end{equation}
\begin{equation}
t_3 = \frac{R_{blr}}{c} \cdot (1 + \cos(\theta + \epsilon))
\end{equation}
\begin{equation}
t_4 = \frac{R_{blr}}{c} \cdot (1 + \cos(\theta - \epsilon))
\end{equation}

This prediction could later be compared to the MACHO observations.  A value of
$\epsilon$ was not inserted until the project had produced results, so that the
mean calculated value of $\epsilon$ could be inserted into the simulation for
comparison with the results.

In the data analysis it was quickly noticed that one of the 59 MACHO quasars -
MACHO 42.860.123 - was very poorly observed by the survey, totalling only 50
observations over the programme's seven-year lifetime.  The V and R data for
the remaining objects were processed in IDL to remove any 10-$\sigma$ data
points -- 5-$\sigma$ would seem a more appropriate value at first glance, but
inspection of the data showed relevant brightness peaks between 5- and
10-$\sigma$.  At the same time all null entries (nights with no observation)
were also removed.  The data were then interpolated to give a uniformly-spaced
brightness record over time, with the number of bins to interpolate over equal
to the number of observations.  The case was made for using a fixed number of
bins for all objects, eg 1000, but for some data records as few as 250
observations were made and so such an interpolation would serve only to
reinforce any remaining spurious points.  By the employed method, a small
amount of smoothing of the data was introduced.  The data then had their
timescales corrected for cosmological redshift to ensure that all plots
produced and any structure inferred were on proper timescales.

In the beginning of the project, to understand the behaviour of the data and
appreciate the complexities and difficulties of the investigation, a list was
compiled of the most highly-sampled ($> 600$ observations) quasars, yielding 30
objects.  Each of these objects had their light curves and autocorrelations
examined and a key feature presented itself immediately; the data showed in
some cases a long-timescale ($\sim 1000$ proper days), large-amplitude ($\sim
0.8$ magnitude) variation that dominated the initial autocorrelation function.
Since this variation was seen only in some quasars (and is on a longer
timescale than the predicted reverberation), this signal was removed by
applying a 300-day running boxcar smooth algorithm over the data, before
subtracting this smoothed data (which would now be low-frequency variation
dominated) from the actual brightness record.  The timescale for smoothing was
determined by examination of the brightness profiles of the objects, which
showed deviations from this
long-timescale signature on timescales below 300 days.  Autocorrelation
analysis of the uncorrected data also found the first autocorrelation minimum
to lie before 300 days.  Further, if the central brightness pulse has a
duration beyond 300 days it will be extremely difficult (if even possible) to
resolve the delay to the first reverberation peak, which is expected to be at a
maximum lag time of several hundred days.  The new autocorrelation calculations
for the long-term variability-corrected data showed from visual inspection the
autocorrelation patterns expected from reverberation.  However, as is described
in section 2, the presence of autocorrelation peaks alone do not necessarily
constitute a detection of reverberation.

To determine the reality of the autocorrelation peaks, a routine was written to
identify the ten largest-amplitude brightening and fading events in each
low-pass filtered brightness record and produce the mean brightness profiles
following them - a mean brightening profile and a mean dimming profile.
Inspection of these brightness profiles demonstrated that any spurious
autocorrelation peaks were a small effect, easily removed by ignoring any peaks
within one hundred days of a larger-amplitude peak.  The justification for this
is simple - if two reverberations occur within one hundred days of each other
(which has a low probability of occurence when one considers the required
quasar orientation), resolving them within the brightness record or
autocorrelation will become difficult, particularly for the most
poorly-sampled, low-redshift objects with a time resolution of $\sim 10$ days
and given that the average half-width of a central brightness peak was found to
be approximately 70 days.  There is also the consideration of the shape of the
reverberation signal from each elliptical surface, which has not yet been
calculated but may have differing asymmetry for each of the four surfaces.

Finally it was noticed that the brightening and dimming profiles do not
perfectly match and in some cases seem to show structure of opposing sign on
the same timescales - if a brightening event was usually followed by another
brightening event at a time lag $t$, a dimming event was sometimes also followed
by a brightening event at a lag of $t$.  The fact that this
effect is not seen in every object makes it difficult to interpret or even
identify as a real effect.  The result of this is that the mean brightness
profile does not always perfectly match the autocorrelation, but usually
demonstrates the same approximate shape.  These arguments are not quantifiable
and so are discarded from future investigation, which will focus on the results
of an automated analysis.  At this point it was also noted that the
highest-redshift quasar in the survey, MACHO 208.15799.1085 at a redshift of
2.77, contained only $\sim 700$ proper days' worth
of observations.  It was felt that this would not be sufficient time to
observe a significant number of reverberation events, which were observed to
occur on timescales of order $\sim 500$ proper days, and so the object was
discarded from further analysis.  Fig. 2 gives one example of the
manually-produced light curves, its autocorrelation and the effect of long-term
variability.

\begin{figure}
\centering
\includegraphics[width=\textwidth]{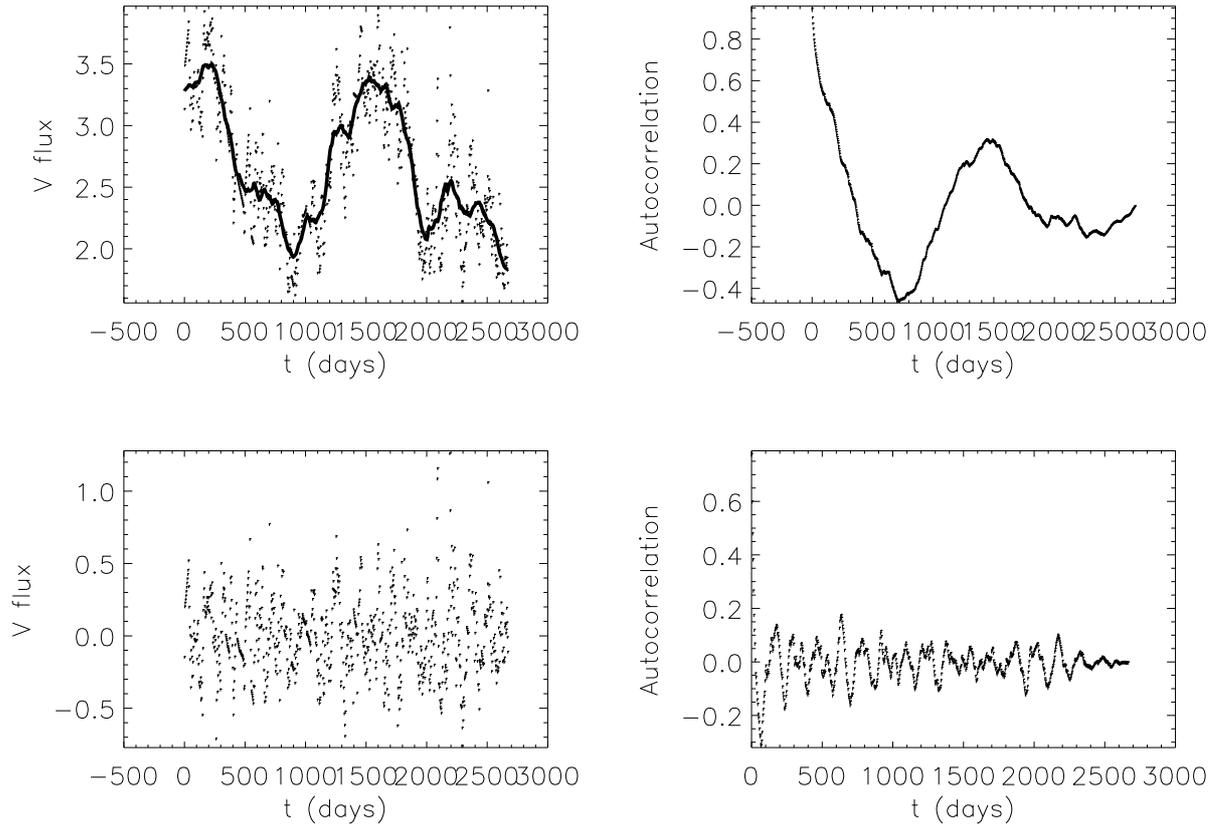}
\caption[Manually-produced plots]{\textbf{Plots produced in manual
inspection of the data for one quasar.}  Top left is the raw data, top right
its autocorrelation.  On the bottom left is the data remaining after
subtraction of long-term variability and bottom right is the corrected data's
autocorrelation function.}
\end{figure}

In the automatically-processed, long-term variability-corrected data,
reverberation patterns were recognised in the autocorrelation by an IDL routine
designed to smooth the autocorrelation on a timescale of 50 days and
identify all positive peaks with lags less than 1000 days.  If two or more
peaks were identified within 100 days of each other, the more prominent peak
was returned.  As a result, we estimate the accuracy of any reverberation
signatures to be approximately 50 days.  Therefore a 50-day smooth
will reduce the number of peaks for the program to sort through while not
detracting from the results, simultaneously removing any double-peaks caused by
dips in autocorrelation caused by unrelated phenomena.  The 50-day resolution
simply reflects the fact that the occurence of two peaks within one hundred
days can not be identified with two separate reverberations.

This technique was also applied to a 'Brownian noise' or simple red noise
simulation with a spectral slope of -2 for comparison.  This was generated
using a random number generator in the following way.  First an array (referred
to as R) of N random numbers was produced and the mean value subtracted from
each number in said array.  A new, blank array (S) of length N+1 was created,
with values then appended as follows:
\begin{equation}
\mbox{if  } R[i] > 0 \mbox{ then } S[i + 1] = S[i] + 1
\end{equation}
\begin{equation}
\mbox{if  } R[i] < 0 \mbox{ then } S[i + 1] = S[i] - 1
\end{equation}

The seeds for the random number generator were taken as the raw data for each
quasar in each filter, so that 114 different simulations were produced to aid
in later comparison with results.
\cite{tk} outline a method for producing more rigorous red noise
simulations of different spectral slopes but to allow more time to focus on the
observations and their interpretation, only this simplified calculation was
performed.

Basic white noise simulations were also produced but produced none of the
predicted autocorrelation structure and so were only followed through the early
data processing stages.  An example white noise autocorrelation is given in
Fig. 3 with a quasar autocorrelation curve overplotted.  Fig. 4 shows the
light curve and autocorrelation function for a star observed by the MACHO
programme compared to one of the quasars to demonstrate the difference in
autocorrelation function and demonstrate that the observed structure is a
feature of the quasars themselves and not induced by observational effects.

\begin{figure}
\includegraphics[width=\textwidth]{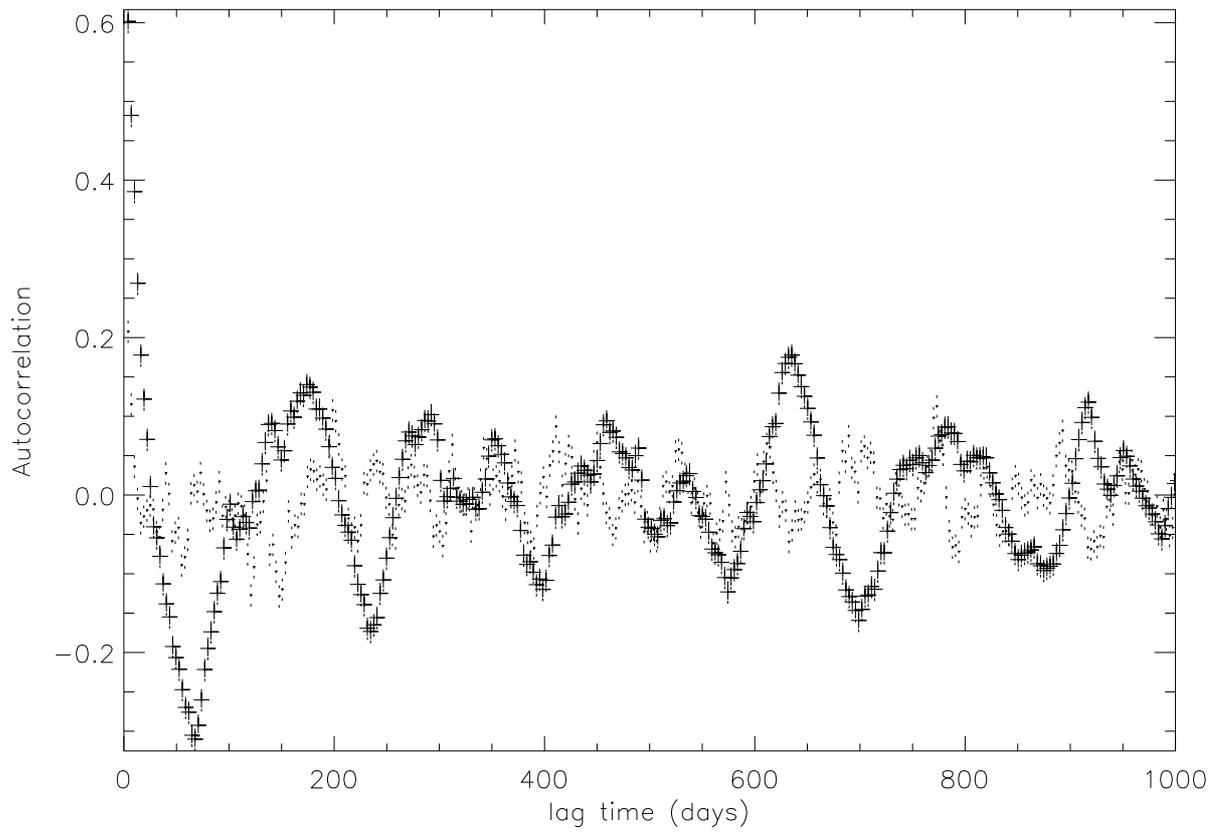}
\caption[Comparison of quasar data to white noise]{\textbf{Quasar (+) and white
noise (.) autocorrelation functions}}
\end{figure}

\begin{figure}
\includegraphics[width=\textwidth]{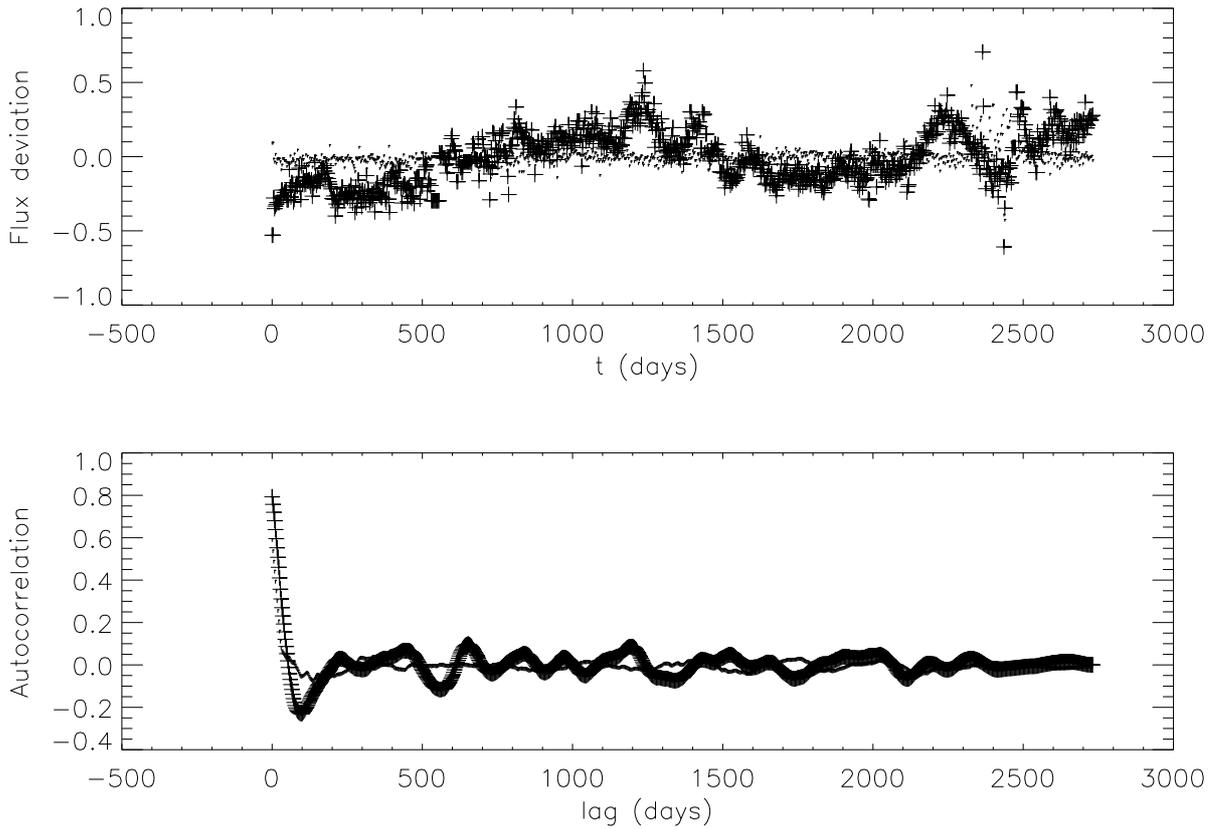}
\caption[Comparison of quasar and star data]{\textbf{Top: Quasar (+) and star
data (.)} demonstrating that the long-term variability observed in the quasars
is not an observational effect.  \textbf{Bottom: Quasar (bold) and star
autocorrelation functions} after 300-day boxcar smooths have been applied,
showing that the autocorrelation structure observed in quasars is intrinsic to
the quasars themselves}
\end{figure}

\section{Results}
Once the data were processed and autocorrelation peaks identified,
the positions of the peaks were then used to calculate the radial distance from
central source to outflow region ($R_{blr}$), the inclination angle of the
observer's line of sight from the accretion disc plane ($\theta$) and a factor
termed the 'internal structure variable' which is the angle from the accretion
disc plane to the reverberating outflow regions ($\epsilon$).  These are
calculated from equations (2-5).  These can then produce
\begin{equation}
(4) + (5) + (6) + (7) = t_1 + t_2 + t_3 + t_4 = 4 \cdot \frac{R_{blr}}{c}
\end{equation}
or $R_{blr} = c \cdot \langle t \rangle$
\begin{equation}
(6) - (5) = t_3 - t_2 = 2 \cdot \frac{R_{blr}}{c} \cdot \cos(\theta + \epsilon)
\end{equation}
or $\theta + \epsilon = \arccos(\frac{t_3 - t_2}{2 \cdot \langle t \rangle})$
\begin{equation}
(7) - (4) = t_4 - t_1 = 2 \cdot \frac{R_{blr}}{c} \cdot \cos(\theta - \epsilon)
\end{equation}
or $\theta - \epsilon = \arccos(\frac{t_4 - t_1}{2 \cdot \langle t \rangle})$

From (11) and (12) we then obtain
\begin{equation}
\theta = \frac{\arccos(\frac{t_3 - t_2}{2 \cdot \langle t \rangle}) + \arccos(\frac{t_4 - t_1}{2 \cdot \langle t \rangle})}{2}
\end{equation}
\begin{equation}
\epsilon = \frac{\arccos(\frac{t_3 - t_2}{2 \cdot \langle t \rangle}) - \arccos(\frac{t_4 - t_1}{2 \cdot \langle t \rangle})}{2}
\end{equation}

Consideration of the geometry involved also shows that three special cases must
be considered.  First is the case where $\theta = 90^o - \epsilon$.  In this
case, peaks two and three are observed at the same time and so only three
distinct brightness pulses are seen.  In analysing the data, the simplifying
assumption was made that all 3-peak autocorrelation signatures were caused by
this effect.  Another case to be considered is when $\theta = 90^o$, in which
cases pulses 1 and 2 are indistinguishable, as are pulses 3 and 4.  Again,
the assumption was made that all two-peak reverberation patterns were caused by
this orientation effect.  Notice that in these cases the equation for $R_{blr}$
is not modified, but the equations for $\epsilon$ in the 3-peak and 2-peak
cases become
\begin{equation}
\epsilon = \frac{\arcsin(\frac{t_3 - t_1}{2 \cdot \langle t \rangle})}{2}
\end{equation}
\begin{equation}
\epsilon = \frac{\arcsin(\frac{t_2 - t_1}{2 \cdot \langle t \rangle})}{2}
\end{equation}
respectively.  Finally there is the case of zero inclination angle, where peaks
one and two arrive at the same time as the initial brightening event, followed
by peaks three and four which then arrive simultaneously.  In this situation
no information about the internal structure variable can be extracted.
Thankfully, no such events were observed.  The results calculated for the 57
studied quasars, from equations (10) and (13-16), are presented in Table 1.
The redshifts presented in Table 1 are as given in \cite{mg}.  Additionally a
column has been included showing $R_{blr}$ in multiples of the minimum measured
$R_{blr}$ for comparison with the relative luminosities of the quasars.

This yields an average $R_{blr}$ of 544 light days with an RMS of 74 light days
and an average $\epsilon$ of $11.87^o$ with an RMS of $2.03^o$.  A plot of
$t_1$, $t_2$, $t_3$, $t_4$ as a function of $\theta$ for the 57 objects was
then compared to the predictions of a simulation using the calculated average
$\epsilon$ from these observations, for the data from both filters.  This
entire process was then repeated for the red noise simulations, of which 114
were produced such that the resulting plots could be directly compared.  The
simulations produced by the R data were treated as R data while the simulations
produced by V data were treated as V data, to ensure the simulations were
treated as similarly to the actual data as possible.  The process was then
repeated for 10 stars in the same field as the example quasar that was studied
in detail, MACHO 13.5962.237.  Notice that while the
quasar data show trends very close to the model, the star data exhibit much
stronger deviations (see Fig. 5).

\begin{deluxetable}{lcccccccccr}
\tablecaption{Properties of the MACHO quasars.  z is the quasar's redshift,
n(V) and n(R) the number of V and R band observations respectively, V and R
respectively are the average V and R magnitudes, $R_{blr}$ the calculated
radial distance from the central source to the reverberating regions, $\theta$
the calculated inclination angle and $\epsilon$ the calculated internal
structure variable.  $R_{rel}$ is the calculated $R_{blr}$ in units of the
smallest $R_{blr}$ found and $L_{rel}$ is the luminosity relative again to the
minimum calculated luminosity.  The last five columns are each an average of
the values calculated separately from the V and R data.}
\tablehead{\colhead{MACHO ID} & \colhead{z} & \colhead{n(V)} & \colhead{V} & \colhead{n(R)} & \colhead{R} & \colhead{$R_{blr}$} & \colhead{$\theta$} & \colhead{$\epsilon$} & \colhead{$R_{rel}$} & \colhead{$L_{rel}$}\\
\colhead{} & \colhead{} & \colhead{} & \colhead{} & \colhead{} & \colhead{} & \colhead{(ld)} & \colhead{($^o$)} & \colhead{($^o$)} & \colhead{} & \colhead{}}
\startdata
2.5873.82 &    0.46 &        959 &      17.44 & 1028 & 17.00 & 608 &  71 & 9.5  & 1.67 & 37\\
5.4892.1971 &    1.58 &        958 &      18.46 & 938 & 18.12 & 560 &  70 & 12.0 & 1.53 & 54\\
6.6572.268 &    1.81 &        988 &      18.33 & 1011 & 18.08 & 578 &  71 & 12.5 & 1.58 & 79\\
9.4641.568 &    1.18 &        973 &      19.20 & 950 & 18.90 & 697 &  72 & 11.0 & 1.91 & 24\\
9.4882.332 &    0.32 &        995 &      18.85 & 966 & 18.51 & 589 &  64 & 12.0 & 1.61 & 6\\
9.5239.505 &    1.30 &        968 &      19.19 & 1007 & 18.81 & 579 &  64 & 12.0 & 1.59 & 28\\
9.5484.258 &    2.32 &        990 &      18.61 & 396 & 18.30 & 481 &  83 & 10.5 & 1.32 & 78\\
11.8988.1350 &    0.33 &        969 &      19.55 & 978 & 19.23 & 541 &  67 & 14.0 & 1.48 & 3\\
13.5717.178 &    1.66 &        915 &      18.57 & 509 & 18.20 & 572 &  59 & 13.0 & 1.57 & 62\\
13.6805.324 &    1.72 &        952 &      19.02 & 931 & 18.70 & 594 &  85 & 9.5 & 1.63 & 41\\
13.6808.521 &    1.64 &        928 &      19.04 & 397 & 18.74 & 510 &  81 & 11.0 & 1.40 & 38\\
17.2227.488 &    0.28 &        445 &      18.89 & 439 & 18.58 & 608 &  82 & 8.0 & 1.67 & 4\\
17.3197.1182 &    0.90 &        431 &      18.91 & 187 & 18.59 & 567 &  73 & 16.0 & 1.55 & 24\\
20.4678.600 &    2.22 &        348 &      20.11 & 356 & 19.87 & 439 &  68 & 14.0 & 1.20 & 18\\
22.4990.462 &    1.56 &        542 &      19.94 & 519 & 19.50 & 556 &  64 & 12.5 & 1.52 & 17\\
22.5595.1333 &    1.15 &        568 &      18.60 & 239 & 18.30 & 565 &  73 & 9.0 & 1.54 & 41\\
25.3469.117 &    0.38 &        373 &      18.09 & 363 & 17.82 & 558 &  65 & 12.0 & 1.53 & 15\\
25.3712.72 &    2.17 &        369 &      18.62 & 365 & 18.30 & 517 &  72 & 12.0 & 1.42 & 73\\
30.11301.499 &    0.46 &        297 &      19.46 & 279 & 19.08 & 546 &  68 & 12.5 & 1.50 & 6\\
37.5584.159 &    0.50 &        264 &      19.48 & 258 & 18.81 & 562 &  70 & 12.5 & 1.54 & 7\\
48.2620.2719 &    0.26 &        363 &      19.06 & 352 & 18.73 & 605 &  65 & 13.0 & 1.66 & 3\\
52.4565.356 &    2.29 &        257 &      19.17 & 255 & 18.96 & 447 &  85 & 8.0 & 1.22 & 44\\
53.3360.344 &    1.86 &        260 &      19.30 & 251 & 19.05 & 496 &  67 & 10.0 & 1.36 & 33\\
53.3970.140 &    2.04 &        272 &      18.51 & 105 & 18.24 & 404 &  69 & 13.5 & 1.11 & 75\\
58.5903.69 &    2.24 &        249 &      18.24 & 322 & 17.97 & 491 &  80 & 8.5 & 1.35 & 104\\
58.6272.729 &    1.53 &        327 &      20.01 & 129 & 19.61 & 518 &  70 & 12.5 & 1.42 & 15\\
59.6398.185 &    1.64 &        279 &      19.37 & 291 & 19.01 & 539 &  83 & 9.5 & 1.48 & 29\\
61.8072.358 &    1.65 &        383 &      19.33 & 219 & 19.05 & 471 &  67 & 16.0 & 1.29 & 29\\
61.8199.302 &    1.79 &        389 &      18.94 & 361 & 18.68 & 475 &  69 & 14.0 & 1.30 & 40\\
63.6643.393 &    0.47 &        243 &      19.71 & 243 & 19.29 & 536 &  67 & 11.0 & 1.47 & 5\\
63.7365.151 &    0.65 &        250 &      18.74 & 243 & 18.40 & 625 &  83 & 10.5 & 1.71 & 19\\
64.8088.215 &    1.95 &        255 &      18.98 & 240 & 18.73 & 464 &  67 & 17.5 & 1.27 & 46\\
64.8092.454 &    2.03 &        242 &      20.14 & 238 & 19.94 & 485 &  73 & 10.5 & 1.33 & 16\\
68.10972.36 &    1.01 &        267 &      16.63 & 245 & 16.34 & 670 &  84 & 10.0 & 1.84 & 216\\
75.13376.66 &    1.07 &        241 &      18.63 & 220 & 18.37 & 561 &  67 & 10.0 & 1.54 & 36\\
77.7551.3853 &    0.85 &        1328 &      19.84 & 1421 & 19.61 & 489 &  69 & 14.0 & 1.34 & 9\\
78.5855.788 &    0.63 &        1457 &      18.64 & 723 & 18.42 & 491 &  77 & 12.0 & 1.35 & 19\\
206.16653.987 &     1.05 &        741 &      19.56 & 581 & 19.28 & 486 &  69 & 14.5 & 1.33 & 15\\
206.17052.388  &    2.15 &        803 &      18.91 & 781 & 18.68 & 365 &  37 & 13.5 & 1.00 & 53\\
207.16310.1050   &  1.47 &        841 &      19.17 & 885 & 18.85 & 530 &  90 & 8.5 & 1.45 & 31\\
207.16316.446 &    0.56 &        809 &      18.63 & 880 & 18.44 & 590 &  66 & 12.5 & 1.62 & 16\\
208.15920.619 &    0.91 &        836 &      19.34 & 759 & 19.17 & 621 &  70 & 13.0 & 1.70 & 15\\
208.16034.100 &    0.49 &        875 &      18.10 & 259 & 17.81 & 742 &  82 & 9.0 & 2.03 & 23\\
211.16703.311  &    2.18 &        733 &      18.91 & 760 & 18.56 & 374 &  90 & 6.5 & 1.02 & 57\\
211.16765.212 &      2.13 &        791 &      18.16 & 232 & 17.87 & 447 &  76 & 13.0 & 1.22 & 108\\
1.4418.1930 &    0.53 &        960 &      19.61 & 340 & 19.42 & 577 &  64 & 11.5 & 1.58 & 6\\
1.4537.1642 &    0.61 &        1107 &      19.31 & 367 & 19.15 & 635 &  79 & 9.0 & 1.74 & 9\\
5.4643.149 &    0.17 &        936 &      17.48 & 943 & 17.15 & 699 &  80 & 9.0 & 1.92 & 8\\
6.7059.207 &    0.15 &        977 &      17.88 & 392 & 17.41 & 613 &  83 & 11.5 & 1.68 & 4\\
13.5962.237 &    0.17 &        879 &      18.95 & 899 & 18.47 & 524 &  66 & 12.5 & 1.44 & 2\\
14.8249.74 &    0.22 &        861 &      18.90 & 444 & 18.60 & 579 &  69 & 13.5 & 1.59 & 3\\
28.11400.609 &    0.44 &        313 &      19.61 & 321 & 19.31 & 504 &  70 & 13.5 & 1.38 & 5\\
53.3725.29 &    0.06 &        266 &      17.64 & 249 & 17.20 & 553 &  69 & 14.0 & 1.52 & 1\\
68.10968.235 &    0.39 &        243 &      19.92 & 261 & 19.40 & 566 &  83 & 12.0 & 1.55 & 3\\
69.12549.21 &    0.14 &        253 &      16.92 & 244 & 16.50 & 497 &  68 & 12.5 & 1.36 & 9\\
70.11469.82 &    0.08 &        243 &      18.25 & 241 & 17.59 & 544 &  79 & 10.5 & 1.49 & 1\\
82.8403.551 &    0.15 &        836 &      18.89 & 857 & 18.55 & 556 &  56 & 12.0 & 1.52 & 1
\enddata
\end{deluxetable}

\begin{figure}
\centering
\subfigure[Quasar R (left) and V (right) data]{\includegraphics[width=\textwidth]{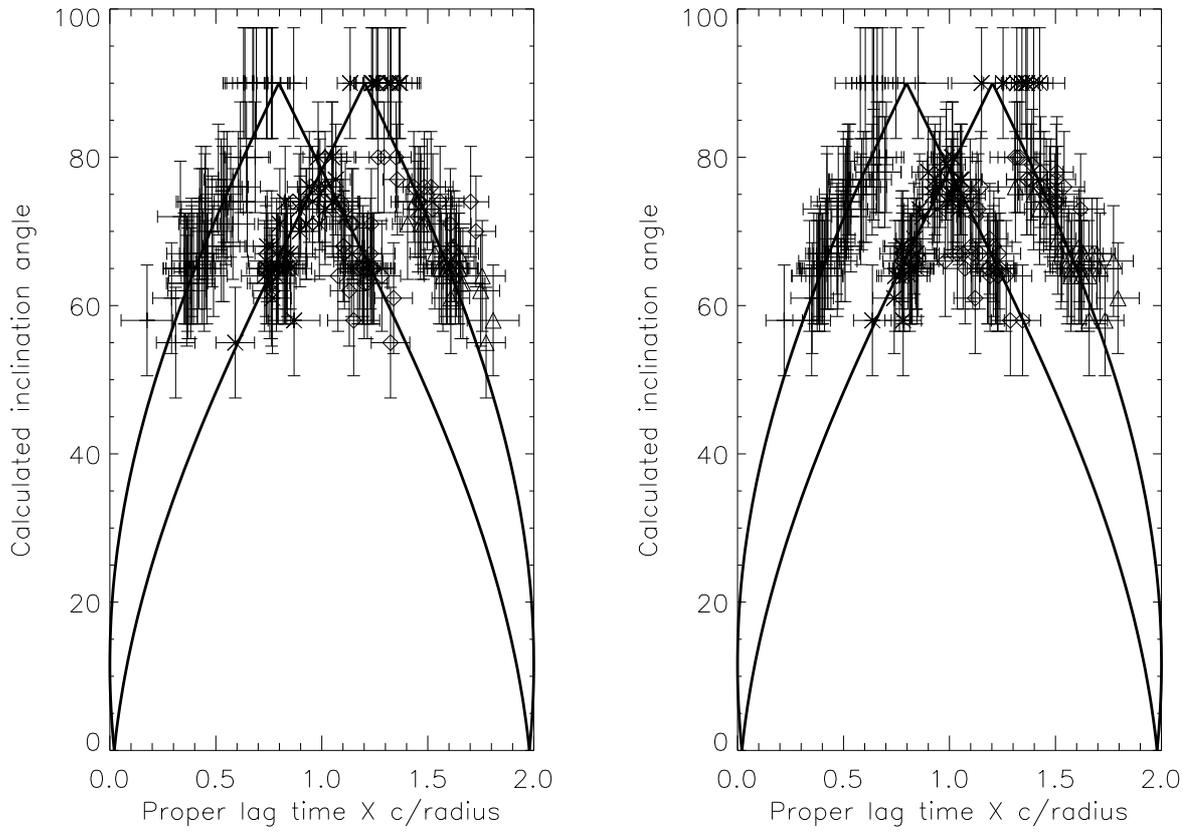}}
\subfigure[Star data fit to model]{\includegraphics[width=0.6\textwidth]{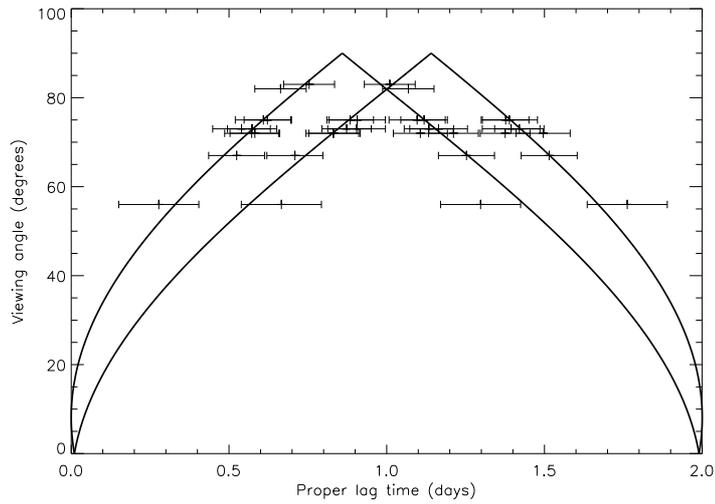}}
\caption[Fit of data to model]{\textbf{Autocorrelation peak lags normalised to
calculated $R_{blr}$ vs calculated inclination angle.}  The solid lines
represent the model's predictions.}
\end{figure}

The red-noise simulation process described in section 5 was repeated ten times
(so that in total 1140 individual red noise simulations were created),
so that a statistic on the expected bulk properties of red noise sources could
be gathered. It was found that on average, $18.5 \%$ of pure red-noise sources
had no autocorrelation peaks detected by the program used to survey the MACHO
quasars, with an RMS of $\pm 6.3 \%$.  Since all 57 studied quasars show
relevant autocorrelation structure, it may be concluded that the observed
structure is a $9 \sigma$ departure from the red-noise distribution, indicating
a greater than $99 \%$ certainty that the effect is not caused by red-noise of
this type.

With an error of $\pm 50$ days in the determination of each lag time, an error
of $\pm 35$ days is present in the calculation of $t_2 - t_1$, $t_3 - t_2$,
etc.  This also yields an error of $\pm 25$ days in the calculated radius of
the outflow region.  If the error in $\cos(\theta \pm \epsilon)$ is small, ie
$\sigma(R_{blr}) \ll R_{blr}$, the small angle approximation may be applied in
calculating the error in radians, before converting into degrees.  Using the
average value of $R_{blr} = 544 ld$ and therefore an error in $\cos(\theta \pm
\epsilon)$ of $\pm 0.0643$, the error in $\theta \pm \epsilon$ is found to be
$\pm 3.69^o$ - sufficiently low to justify the small angle approximation.  The
errors in $\theta$ and $\epsilon$ are therefore both equal to $\pm 2.61^o$.
The errors in the calculated mean values for $R_{blr}$ and $\epsilon$ are then
$\pm 3.3$ light days and $\pm 0.35^o$ respectively.  The 100-day resolution
limit for reverberation peaks also restricted the value of $\theta$ to greater
than $50^o$ - at lower inclinations the program was unable to resolve
neighbouring peaks as they would lie within 100 days of each other for a quasar
with $R_{blr} \sim 544$ light days.  For this reason, the calculated
inclination angles are inherently uncertain.  The contribution to the RMS of
$\epsilon$ is surprisingly small for the two- and three-peak objects.

The availability of data in two filters enables an additional statistic to be
calculated - the agreement of the R and V calculated values.  For one quasar,
MACHO 206.17057.388, one autocorrelation peak was found in the V data while
three were found in the R.  Inspection of the data themselves showed that the
R data was much more complete and so the V result for this object was
discarded.  The mean deviation of the R or V data from their average was found
to be 44 days, giving an error on the mean due to the R/V agreement of $\pm
4.1$ days.  The mean deviation of the calculated $\theta$ due to the R/V
agreement is $7.1^o$ and that for $\epsilon$ is $2.04^o$, yielding an error in
$\epsilon$ due to the difference in R and V autocorrelation functions of $\pm
0.19^o$.

The RMS deviation of the observed lag times from those predicted by a
model with internal structure variable $\epsilon = 11.87^o$ is found to be
$4.29^o$, which is acceptable given a systematic error of $\pm 2.61^o$ and an
RMS in the calculated $\epsilon$ of $2.03^o$.  Combining the two errors on each
of the mean $R_{blr}$ and $\epsilon$ yields total errors on their means of
$\pm 5.26$ light days and $0.40^o$ respectively.

The interpretation of $\epsilon$ is not entirely clear - it is presented in 
\cite{S05} as the angle made by the luminous regions of the outflowing winds
from the accretion disc.  However, as is demonstrated in Fig. 6a, $\epsilon$
may in fact represent the projection angle of the shadow of the accretion disc
onto the outflowing winds if the central luminosity source occupies a region
with a radius lower than the disc thickness, thus giving by geometry the ratio
of the inner accretion disc radius to the disc thickness.  If the central
source is extended above and below the accretion disc, however, then $\epsilon$
may represent, as shown in Fig. 6b, the ratio of disc thickness to $R_{blr}$.
There may also be a combination of effects at work.

\begin{figure}
\subfigure[Interpretation of $\epsilon$ for a central source much smaller than
the thickness of the accretion disc]{\includegraphics[width=\textwidth]{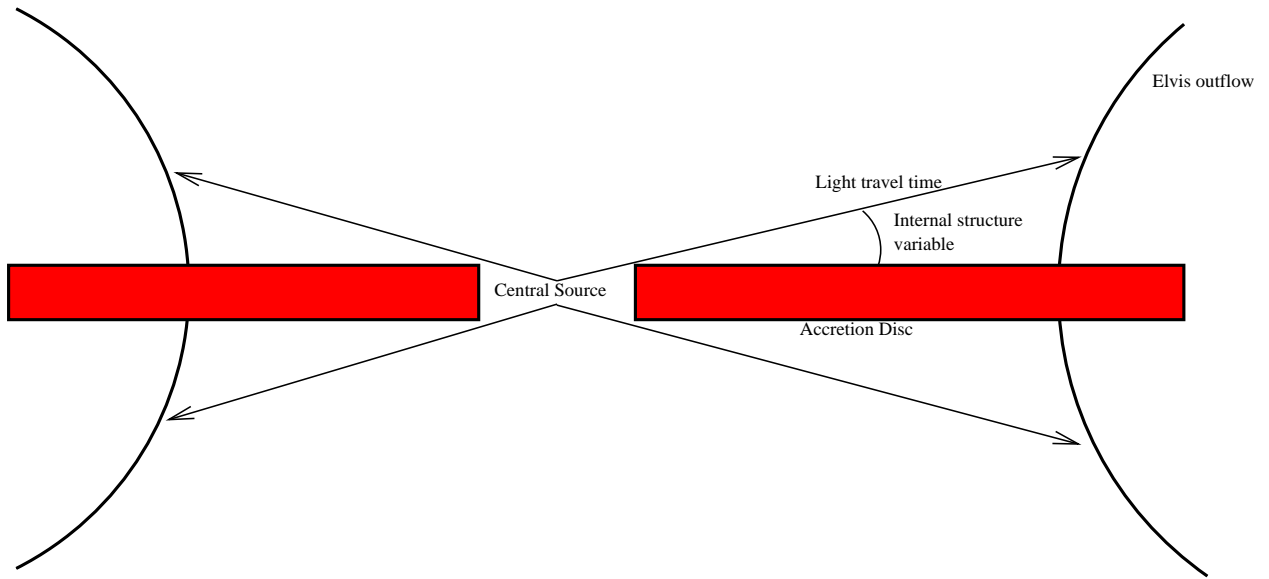}}
\subfigure[Interpretation of $\epsilon$ if the central source extends above
and below the accretion disc]{\includegraphics[width=\textwidth]{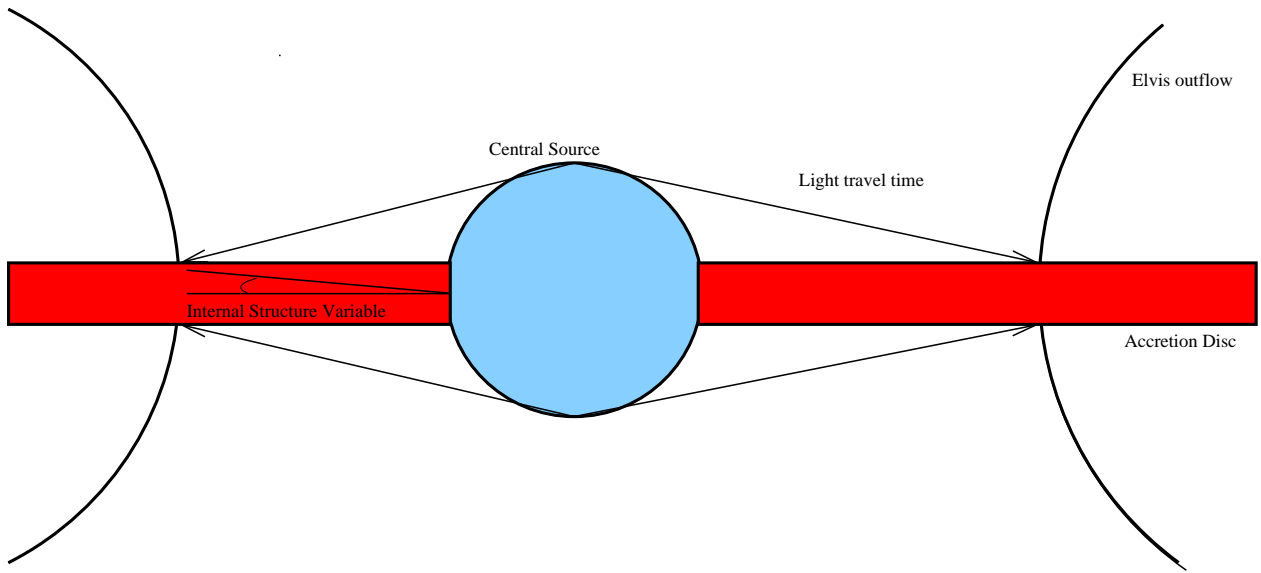}}
\caption[Interpretations of $\epsilon$]{\textbf{Interpretations of $\epsilon$}}
\end{figure}

An accurate way of discriminating between these two interpretations may be
found by analysing the mean reverberation profiles from the four outflow
surfaces and producing a relation between the timing of the central
reverberation peak and that of the onset of reverberation by a specific outflow
surface.  The two figures above demonstrate how the timing of reverberation
onset is interpreted in the two cases of compact vs extended central source
but to discriminate, models must be produced of the expected reverberation
profiles for these two cases.  Results from \cite{slr06} would predict the
former, as indeed would \cite{cs} and \cite{VA07} and so figure 6a would seem
favourable, though analysis of the reverberation profiles will enable a more
concrete determination of the meaning of $\epsilon$.  Models may then be
produced of the expected brightness patterns for differing outflow geometries
and compared to the observed light curves to determine the curvature and
projection angles of the outflows.

The calculated luminosities may now be compared to the calculated $R_{blr}$ for
this sample to determine whether the \cite{ks} or \cite{b} relations hold for
quasars.  Fig. 7 demonstrates that there is absolutely no correlation found
for this group of objects.  This is not entirely unsurprising as even for
nearby AGN both \cite{ks} and \cite{b} had enormous scatters in their results,
though admittedly not as large as those given here.  One important thing to
note is that given a predicted factor 200 range in luminosity, one might take
\cite{b} as a lower limit on the expected spread, giving a predicted range of
$R_{blr}$ of approximately a factor 15.  The sensitivity range of this project
in $R_{blr}$ is a function of the inclination angle of the quasar but in theory
the $R_{blr}$ detectable at zero inclination would be 500 light days with a
minimum detectable $R_{blr}$ at this inclination of 50 light days.  This
factor 10 possible spread of course obtains for any inclination angle giving
possibly an infinite spread of $R_{blr}$.  For the calculated RMS of 77 light
days, the standard deviation is 10.2 light days, 2\% of the calculated mean.
The probability of NOT finding a factor 15 range of $R_{blr}$ if it exists is
therefore less than 1\%.  Therefore it can be stated to high confidence that
this result demonstrates that the \cite{ks} and \cite{b} results do not hold
for quasars.  The failure of \cite{ks} for this data set is evident from the
fact that it predicts an even larger spread in $R_{blr}$ than \cite{b} which
is therefore even less statistically probable from our data.  It is possible
that this is a failure of the SED of \cite{SED}, that there are some as-yet
unrecognised absorption effects affecting the observations or that the
assumption of negligable absorption/re-emission times is incorrect but the
more likely explanation would seem to be that no correlation between $R_{blr}$
and luminosity exists for quasars.
\begin{figure}
\includegraphics[width=\textwidth]{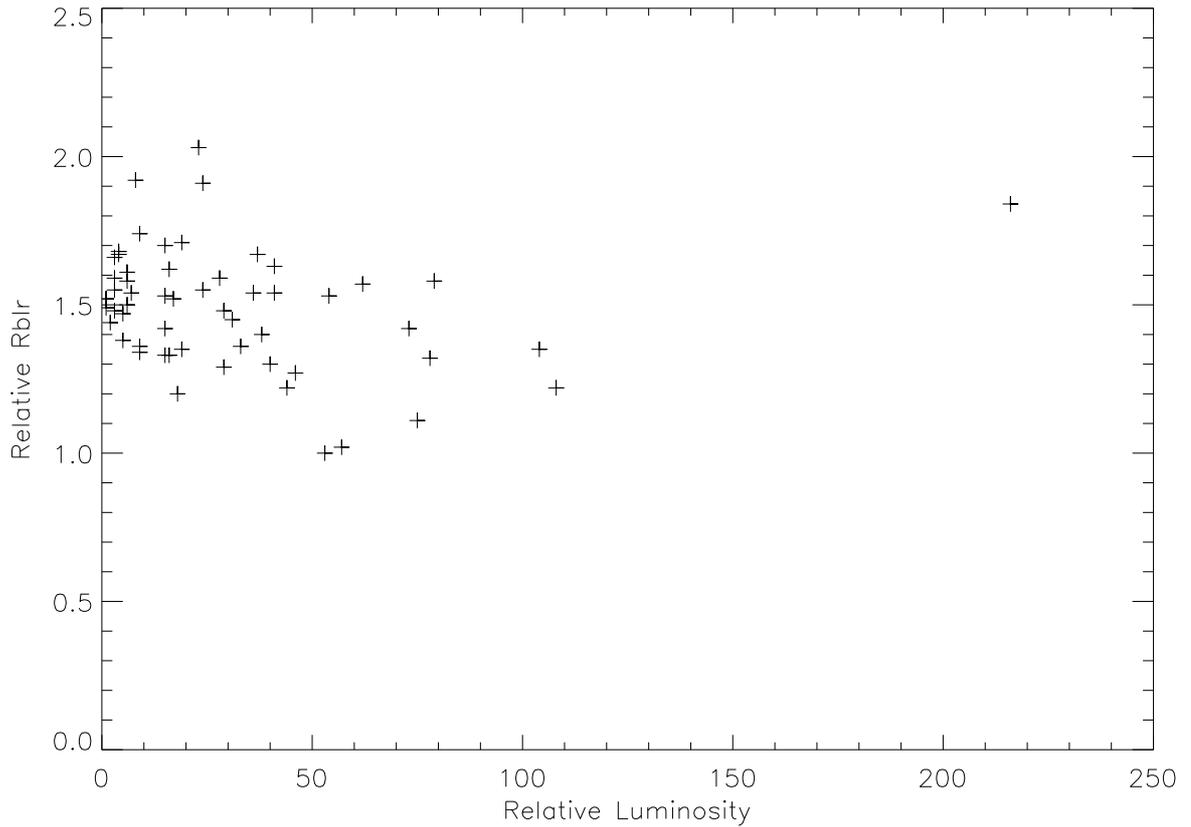}
\caption[Relative luminosities and radii]{\textbf{Relative luminosities and
broad line radii.}  Absolutely no correlation is seen
between $R_{rel}$ (the relative $R_{blr}$) and $L_{rel}$ (the relative
luminosity), demonstrating that the \cite{ks} and \cite{b} results do not
hold for quasars.  The error bars are so small that they are contained within
the data points.}
\end{figure}

A calculated average $R_{blr}$ of 544 light days corresponds to $1.4 \times
10^{18} cm$, compared to the predicted average of $\sim 10^{18} cm$.  The value
found here is in remarkable agreement with those previously calculated,
made even more surprising by the fact that so few results were available upon
which to base a prediction.  Many quasar models predict a large variation in
quasar properties, see for example \cite{wu}, so we conclude that
quasars and perhaps AGN in general are an incredibly homogeneous population.

\section{Conclusions}
Brightness records of 57 quasars taken from the MACHO survey in R
and V colour filters have been analysed to show the presence of autocorrelation
structure consistent with biconic outflowing winds at an average radius of
$544 \pm 5.2$ light days with an RMS of 74 light days.  An internal
structure variable of $11.87 \pm 0.40^o$ was found, with an RMS of $2.9^o$.
The accuracy of the program designed to determine the timing of the
reverberation peaks limited its temporal resolution to 100 days, resulting in
the quoted systematic errors in the mean values calculated.  With
longer-timescale, more regularly sampled data this temporal resolution can be improved
- this may also be achieved with more sophisticated computational techniques
combined with brightness models not available in this project.

The correlations between radius of the broad emission line region and
luminosity found by \cite{ks} and \cite{b} for nearby AGN do not appear to hold
for quasars.  This may be indicative of some time or luminosity evolution of
the function as no redshift-independant correlation is found in this data set.
If there is some relation it is more likely to be time-evolving since any
luminosity dependance would most likely be noticeable in Fig. 7, which it is
not.

The presence of reverberation in 57 of the 57 quasars analysed implies that
the outflowing wind is a universal structure in quasars - a verifiable
result since this structure may be identified in regularly-sampled quasars in
other surveys.  While it is acknowledged that red noise may yet be responsible
for the brightness fluctuations observed, the results are so close to the
initial model's prediction that noise seems an unlikely explanation, especially
given the corroborating evidence for the theory \cite{S05,VA07,p,bp,r}.

The continuum variability of quasars, though well observed, is still not well
understood.  The results of this study would suggest that an understanding of
these fluctuations can only be found by recognising that several physical
processes are at work, of which reverberation is of secondary importance in
many cases.  It does however appear to be universally present in quasars and
possibly in all AGN.  Given that quasars are defined observationally by the
presence of broad, blue-shifted emission lines, of which outflowing winds are
the proposed source, this result is strong support for the \cite{E00} model.

\section{Future Work}
Several phenomena identified in the MACHO quasar light curves remain as yet
unexplained.
\begin{enumerate}
\item
What is the source of the long-term variability of quasars?  Is it a random
noise process or is there some underlying physical interpretation?  It has
been suggested that perhaps a relativistic orbiting source of thermal emission
near the inner accretion disc edge may be the source of such fluctuation.
Modelling of the expected emission from such a source must be undertaken
before such a hypothesis can be tested.
\item
Why is it that the brightness profile following a dimming event sometimes
agrees perfectly with the brightening profile while at other times it is in
perfect disagreement?  Again, is this a real physical process?  Work by
\cite{ge} on stratified wind models presents a situation
where the central object brightening could increase the power of an inner wind,
increasing its optical depth and thus shielding outer winds.  This would result
in negative reverberation.  Further investigation may demonstrate a dependance
of this effect on whether the central variation is a brightening or fading.
\item
What is the mean profile of each reverberation peak?  This profile may yield
information about the geometry of the outflowing wind, thus enabling
constraints to be placed on the physical processes that originate them.
\item
Can quasars be identified by reverberation alone?  Or perhaps by the long-term
variability they exhibit?  With surveys such as Pan-STARRS and LSST on the
horizon, there is growing interest in devloping a purely photometric method of
identifying quasars.
\item
LSST and Pan-STARRS will also produce light curves for thousands of quasars
which can then be analysed in bulk to produce a higher statistical accuracy
for the long-term variability properties of quasars.  It is evident that the
sampling rate will only be sufficient for reverberation mapping to be performed
with LSST and not Pan-STARRS.
\item
Is there a time- or luminosity-evolving relation between $R_{blr}$ and
luminosity?  Comparison of $R_{blr}$, luminosity and redshift may yet shed
light on this question.
\end{enumerate}

\section{Acknowledgements}
I would like to thank Rudy Schild for proposing and supervising this project,
Pavlos Protopapas for his instruction in IDL programming, Tsevi Mazeh for his
advice on the properties and interpretation of the autocorrelation function
and Phil Uttley for discussion and advice on stochastic noise in quasars.
This paper utilizes public domain data originally obtained by the MACHO
Project, whose work was performed under the joint auspices of the U.S.
Department of Energy, National Nuclear Security Administration by the
University of California, Lawrence Livermore National Laboratory under
contract No. W-7405-Eng-48, the National Science Foundation through the
Center for Particle Astrophysics of the University of California under
cooperative agreement AST-8809616, and the Mount Stromlo and Siding Spring
Observatory, part of the Australian National University.


\newpage
\addcontentsline{toc}{section}{References}
\begin{thebibliography}{}
\bibitem{E00} [1] Elvis, M., ApJ, 545, 63 (2000)
\bibitem{ms}  [2] Matthews, T. \& Sandage, A., ApJ, 138, 30 (1963)
\bibitem{wm}  [3] Webb, W. \& Malkan, M., ApJ, 540, 652 (2000)
\bibitem{O}  [4] Ogle, P., PhD Thesis, CalTech (1998)
\bibitem{FEA} [5] Fabian, A. et al., arXiv: 0903.4424 (2009)
\bibitem{p00} [6] Proga, D., ApJ, 538, 684 (2000)
\bibitem{p08} [7] Proga, D., Ostriker, J. \& Kurosawa, R., ApJ, 676, 101 (2008)
\bibitem{ra}  [8] Antonucci, R., Ann. Rev. Ast. \& Ap. 31, 473 (1993)
\bibitem{mp}  [9] Magdis, G. \& Papadakis, I., ASPC, 360, 37 (2006)
\bibitem{ST97} [10] Schild, R. \&  Thomson, D., AJ, 113, 130 (1997)
\bibitem{S05} [11] Schild, R., AJ, 129, 1225 (2005)
\bibitem{VA07}[12] Vakulik, V. et al., MNRAS, 382, 819 (2007)
\bibitem{slr08}[13] Schild, R., Leiter, D. \& Robertson, S., AJ, 135, 947 (2008)
\bibitem{lslp} [14] Lovegrove, J. et al., in preparation (2009)
\bibitem{rl07}  [15] Robertson, S. \& Leiter, D., in New Developments in Black Hole Research (Nova Science Publishers, New York, 2007) pp1-48
\bibitem{rl03}   [16] Robertson, S. \& Leiter, D., ApJ, 596, 203 (2003)
\bibitem{ks} [17] Kaspi, S. et al., ApJ, 629, 61 (2005)
\bibitem{b}   [18] Bentz, M. et al., arXiv: 0812.2283 (2008)
\bibitem{wu}   [19]  Woo, J.-H. \& Urry, C. M., ApJ, 579, 530 (2000)
\bibitem{sll} [20] Schild, R., Leiter, D. \& Lovegrove, J., in preparation (2009)
\bibitem{p}    [21] Pooley, D. et al., ApJ, 661, 19 (2007)
\bibitem{ss73} [22] Shakura, N. \& Sunyaev, R., A\&A, 24, 337 (1973)
\bibitem{bp}  [23] Peterson, B. et al., Ap J, 425, 622 (1994)
\bibitem{r}  [24] Richards, G. et al., ApJ, 610, 671 (2004)
\bibitem{w} [25] Wyithe, S., Webster, R. \& Turner, E., MNRAS, 318, 1120 (2000)
\bibitem{k}     [26] Kochanek, C., ApJ, 605, 58 (2004)
\bibitem{ei}  [27] Eigenbrod, A. et al., A\&A, 490, 933 (2008)
\bibitem{t} [28] Trevese, D. et al., ApJ, 433, 494 (1994)
\bibitem{h96} [29] Hawkins, M., MNRAS, 278, 787 (1996)
\bibitem{h06} [30] Hawkins, M., ASPC, 360, 23 (2006)
\bibitem{cs}  [31] Colley, W. \& Schild, R., ApJ, 594, 97 (2003)
\bibitem{h07} [32] Hawkins, M., A\&A, 462, 581 (2007)
\bibitem{slp} [33] Schild, R., Lovegrove, J. \& Protopapas, P., astro-ph/0902.1160 (2009)
\bibitem{u} [34] Uttley, P. et al., PTPS, 155,170 (2004)
\bibitem{as}  [35] Arevalo, P. et al., MNRAS, 389, 1479 (2008)
\bibitem{gi}  [36] Giveon, U. et al., MNRAS, 306, 637 (1999)
\bibitem{r04} [37] Rengstorf, A. et al., ApJ, 606, 741 (2004)
\bibitem{n}  [38] Netzer, H., MNRAS, 279, 429 (1996)
\bibitem{bot} [39] Botti, I. et al., astro-ph/0805.4664 (2008)
\bibitem{macho} [40] Alcock, C. et al., PASP, 111, 1539 (1999)
\bibitem{mg}  [41] Geha, M. et al., AJ, 125, 1 (2003)
\bibitem{SED} [42] Richards, G. et al., astro-ph/0601558 (2006)
\bibitem{tk} [43] Timmer, J. \& Koenig, M., A\&A, 300, 707 (1995)
\bibitem{slr06} [44] Schild, R., Leiter, D. \& Robertson, S., AJ, 132, 420 (2006)
\bibitem{ge}  [45] Gallagher, S. \& Everett, J., ASPC, 373, 305 (2007)
\end{thebibliography}
\end{document}